\documentclass[a4paper,11pt,draft]{revtex4}
\usepackage{graphicx}
\usepackage[applemac]{inputenc}
\usepackage{t1enc}
\usepackage{lscape}
\usepackage{amsmath}
\usepackage{amssymb}

\usepackage{epsfig}

\begin{document}

\title{Validity of path thermodynamics in reactive systems}

\author{M. Malek Mansour}
\affiliation{
Center for Nonlinear Phenomena and Complex Systems,\\
Universit\'e Libre de Bruxelles CP 231, Campus Plaine,\\
B-1050 Brussels, Belgium}

\author{Alejandro L. Garcia}
\affiliation{Dept. Physics and Astronomy, San Jose State University,
San Jose, California, 95192 USA}

\date{\today}

\begin{abstract}
Path thermodynamic formulation of non-equilibrium reactive systems is considered.  It is shown through simple practical examples that this approach can lead to results that contradict well established thermodynamic properties of such systems. Rigorous mathematical analysis confirming this fact is presented.
\end{abstract}

\maketitle

Stochastic thermodynamics was introduced in the early eighties in the framework of jump Markov processes, using the Gibbs-Shannon definition of entropy $S(X, t) = - k_B \ln P(X, t)$, where $k_B$ is the Boltzmann constant, $X \in \mathbb{Z}^N$ the (random) composition of an $N$-component reactive system ($N < \infty$), and $P(X, t)$ the associated probability distribution at time $t$ \cite{Nico}. Path thermodynamics is a generalization of stochastic thermodynamics. It is based on the concept of "path entropy", an entropy associated with sample paths (trajectories) of a stochastic process. Nowadays the general belief is that path thermodynamics is the ultimate theoretical formalism for physico-chemical systems ranging from macroscopic to nanometer scale \cite{Rev1,Harris,Rev2,Jarz4}. Though, the main difference with the traditional Gibbs-Shannon stochastic thermodynamics is that path thermodynamics leads to the so-called "fluctuation theorem" for the path entropy production (see eq. (\ref{Eq7}) for the precise definition).  This fundamental relation implies  that path entropy production may take negative values in non-equilibrium physico-chemical systems, in apparent contradiction with the second law of thermodynamics (see \cite{MF}). The probability of observing such an event is essentially zero in macroscopic systems \cite{Harris,Rev2}, but not necessarily so in very small reactive systems, like those encountered at the level of biological cells or simulated by molecular dynamics. And, as such, it was at the origin of the impressive resurgence of interest in the path thermodynamics description of reactive systems, generally modeled as a jump Markov process.

Dozens and dozens of articles have appeared over the last decade on this topic, all of them treating path thermodynamics as a fundamental, well-established theory. Yet, we shall prove that the resulting properties will be wrong whenever the reactive system involves more than one elementary reaction leading to the same composition changes. Needless to say, this is the case for the great majority of the reactive models considered in these papers.  Our results will thus have significant impact on future work in this field.

In this paper we shall be concerned mainly with reactive systems in contact with constant (time-independent) reservoirs and modeled as a jump Markov process.  It is organized as follows: First, we present simple examples to illustrate the core of the problem so that readers not familiar with the subject can easily comprehend the issue raised in this paper.  Next, after presenting a general background summary, the relevant mathematical aspects are rigorously derived in a series of proofs. Finally, we close with some  concluding remarks.

\section{Illustrative examples}
Consider a perfectly homogeneous dilute reactive system, as can be produced experimentally in a  "continuously stirred tank reactor" (CSTR).  The CSTR is a particularly useful device that maintains a reactive system out of equilibrium by fixing the concentration of some reactants \cite{Vidal,Adv}.  In principle, by recording the step-by-step evolution of the system, a large number of sample paths can be collected for detailed analysis.  In practice, such data are usually obtained by numerical simulation, using a well-established algorithm introduced some decades ago by Gillespie \cite{Gil1}.  Suppose now that two of the reactions, say $R_1$ and $R_2$, are of the type $X \mathop{\rightleftharpoons} Y$ and  $ X + Y \mathop{\rightleftharpoons} 2 Y.$     Starting from an arbitrary state $(X, Y \ne 0)$, both reactions lead to the same change, that is  $(X, Y) \rightarrow (X-1, Y+1)$ (forward) or $(X, Y) \rightarrow (X+1, Y-1)$ (backward).  Consequently, when analyzing the sample paths of this system, there is no criterion that allows us to differentiate $R_1$ from $R_2$.  However, we know from the basic principles of irreversible thermodynamics that the entropy production of reactive systems depends on the characteristics of each \emph{individual} reaction (see, for example, Section 9.5 of ref. \cite{Dilip}).
We thus conclude that in this case the entropy production based on path thermodynamics corresponds actually to a "coarse grained" entropy production which is inconsistent with irreversible thermodynamics results \cite{MF,Kawai,Vdb1}.
Note that a similar observation was reported by Seleznev et al. \cite{Russe} (see also \cite{Kurch} and \cite{DDK}).

It is important to notice that this example is far from being an isolated, exceptional case.  In fact, the  majority of published works dealing with complex behavior in reactive systems involves at least two elementary reactions leading to the same composition change.  Well known examples are the Schl{\"o}gl model and the reversible "Brusselator" \cite{NicPrig}, as well as many of the reactive systems related to the modeling of biological systems \cite{Epst}.

An unexpected consequence of this issue is that the sample paths of a class of one variable reactive systems behave, in a stationary non-equilibrium regime, like those of a system at thermodynamic equilibrium.  For instance, let us consider again the example of a dilute reactive mixture in a CSTR that is now set up in order to maintain constant the concentration of all but one of the reactants,  say $X$.  Suppose that the reactions involving that component $X$ are of the type
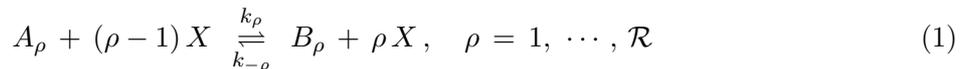
\begin{equation}
\label{Eq1}
A_\rho \, + \,  (\rho - 1) \, X \,\,\, \mathop{\rightleftharpoons}^{k_\rho}_{k_{- \rho}} \,\,\, B_\rho \, + \, \rho \, X \, , \quad \rho \, = \,  1, \, \, \cdots, \, \mathcal{R}
\end{equation}
Unlike the previous example, here \emph{all of the reactions} lead exactly to the same changes, that is,  $X \rightarrow X+1$ (forward) or $X \rightarrow X-1$ (backward). As mentioned above, the sample paths of this system do not permit us to distinguish one reaction from another. They are thus equivalent to sample paths of a system with a single reversible reaction, symbolically represented as $X \mathop{\rightleftarrows} X + 1$. This equivalence in turn implies that the corresponding sample paths are perfectly reversible in the stationary regime, in the sense that any path joining an arbitrary state $S_1$ to another arbitrary state $S_2$ will occur with the same probability as the corresponding reverse path joining $S_2$ to $S_1$ (the mathematical proof will be presented below).   But such time-reversal symmetry concerns only the thermodynamic equilibrium state where, on average, each forward reaction  in (\ref{Eq1}) is exactly balanced by its reverse. We are thus faced with a paradox.  On the one hand, in the stationary regime, the sample paths of this system, under given (time-independent) non-equilibrium constraints, possess time-reversal symmetry, which is the key signature of thermodynamic equilibrium states \cite{DegMaz}.  And, on the other hand, the Gibbs-Shannon entropy production of the same system, under the very same non-equilibrium conditions, proves to be strictly positive, converging to the macroscopic thermodynamic result in the vanishing noise limit \cite{Nico}.

Given the controversial nature of the issues illustrated by the above examples, we now concentrate on the strictly mathematical aspects of the problem.

\section{General background}
Let $\chi(t)$ be a pure $N$-dimensional ($N < \infty$) jump Markov process, that is a Markov process with countable state space ${\cal E} \in \mathbb{Z}^N$  which has all its sample paths constant, except for isolated jumps.  A jump Markov process is entirely characterized by  the so called {\it transition probabilities per unit time}, or  {\it transition rates}, defined as \cite{Gard,VK}
\begin{equation}
\label{Eq2}
W(X \, | \,  X')   =  \lim_{\Delta t \, \downarrow \, 0} \, \frac{1}{\Delta t} \, \Big[ P(X, t + \Delta t \, | \, X', t) \, - \, P(X, t \, | \, X', t) \Big]
\end{equation}
where the limit is assumed to be uniform in time $t$.  The function $P(X, t + \Delta t \, | \, X', t)$ represents the conditional probability to have $\chi(t + \Delta t) = X$, given that $\chi(t) = X'$.
Noticing that $P(X, t \, | \, X', t) = \delta^{Kr}_{X', \, X \,}$,  we deduce that $W(X \, | \,  X') \, \ge \, 0, \, \forall \, X'  \ne  X$. And since the convergence in (\ref{Eq2}) is assumed to be uniform in time,  we have  $\sum_{X}  \, W(X \, | \,  X')  =  0$.  Finally, using the Markovianity of  $\chi(t)$, it can be shown that the conditional probability distribution obeys the so-called {\it Kolmogorov forward equation}, or {\it master equation} \cite{Gard,VK},
\begin{equation}
\label{Eq3}
\frac{d}{d t} \, P(X, \, t \, | \, X_0, \, t_0 )\, = \,  \sum_{X'}  \, W(X \, | \,  X') \, P(X', \, t \, | \, X_0, \, t_0 )
\end{equation}
for $t  >  t_0$, with $\lim_{ t \, \downarrow \, t_0}  P(X,  t  \, | \, X_0,  t_0 ) = \delta^{Kr}_{X, \, X_0}$.
We note that the definition (\ref{Eq2}) implies that
\begin{equation}
\label{Eq4}
P(X, t + \Delta t \, | \, X', \, t)  =  W(X \, | \,  X') \, \Delta t \, + \, {\rm o}(\Delta t) \, , \,\, \forall \, X \, \ne \, X'
\end{equation}

Let us assume that from any state $S \in {\cal E}$, any other state $S' \in {\cal E}$ can be reached by a succession of jumps.   This assumption excludes the existence of absorbing states as well as uni-directional processes, such as "pure birth" or "pure death" processes.  If the state space ${\cal E}$ is finite, then this  assumption also implies that there exists a unique stationary probability distribution $P_s(X)$ that is reached exponentially fast in time from an arbitrary initial state $X_0 \in {\cal E}$.   As far as physico-chemical systems are concerned, it is reasonable to assume that this property remains true even if ${\cal E}$ is not finite.  Here, however, all we need is to assume that $\chi(t)$ remains bounded, almost surely, at least over a finite time interval $[t_0, \, t_f]$.  It can then be proved that the master equation (\ref{Eq3}) has a unique solution in this time interval \cite{Kolmo}. We will show that this unicity theorem, due to Kolmogorov and Feller, effectively restricts the class of reactive systems that can be described by path thermodynamics.

Let us now concentrate on sample paths of $\chi(t)$.  We consider a finite time interval $[t_0, \, t_f]$ that we divide into $n$ sub-intervals, $t_0 \,  < \,   \cdots < \,  t_n \, \equiv t_f$, with  $n \gg 1$.  We then introduce the (joint) probability distribution  $P({\bf X})$,  where ${\bf X} \equiv \big\{X_0, t_0 \, ; \, X_1, t_1 \, ; \, \cdots \, ; \, X_{n-1}, t_{n-1} \, ; \,  X_f, \, {t_f} \big\}$ represents a sample path of $\chi(t)$ that starts from the state $X_0$ at the initial time $t_0$ and ends up at the state $X_f$ at the final time ${t_f}$.

Similarly, we introduce the "reverse" joint probability distribution $P^{(R)}(\widetilde{\bf X})$,  where $\widetilde{{\bold X}} \equiv \big\{X_f, t_0 \, ; \, X_{n-1}, {t}_1 \, ; \, \cdots \, ; \,  X_1, {t}_{n-1} \, ; \,  X_0, {t_f} \big\}$ represents the "reverse sample path" of $\chi(t)$ that starts from the final state $X_f$ at the initial time $t_0$ and ends up at the initial state $X_0$ at the final time ${t_f}$. Note that the superscript "$(R)$" in $P^{(R)}(\widetilde{\mathbf{X}})$ indicates that we are dealing with the probability of the reverse path. In particular, the change of variables ${\bold X} \rightarrow \widetilde{\bold X}$ transforms $P$ to $P^{(R)}$, and vice versa.

Before proceeding further, it is important to notice that,
%the sequence $t_0, t_1, \cdots, t_n$ is \emph{not} the times at which the system undergoes a transition. In fact,
contrary to what is stated by some authors, a conditional probability $P(X_{i+1}, t_{i+1} \,  | \, X_i, t_i)$ {\it does not} correspond to the probability of observing the jump  $X_i \rightarrow X_{i+1}$ at the given instant of time  $t_{i+1}$, the jump may actually occur at any instant of time in the interval $] t_i, \, t_{i+1}]$.  Besides, from the strict mathematical point of view, the statement "observing a jump at a given instant of time" is simply of zero measure \cite{Kolmo}.  Furthermore, it is worthwhile to recall a basic Kolmogorov theorem stating that stochastic processes (including those with continuous realizations) are completely characterized by their family of discretized sample paths, such as ${\bf X}$, provided the associated probability distribution exists and remains invariant under the permutation of the pairs  $(X_i, \, t_i)$ and $(X_j, \, t_j)$, $\forall \, i, j \in [0, \, n]$ \cite{Kolmo,Soong}.

Let us denote respectively by $\Delta S$ and $\Delta_eS$ the entropy variation and the entropy flow along the sample path ${\bf X}$ joining the initial state $X_0$ to the final state $X_f$. Similarly, we denote by $\Delta_iS = \Delta S - \Delta_eS$ the corresponding entropy production, also known as "path entropy production".  The cornerstone of path thermodynamics is the relation
\begin{equation}
\label{Eq5}
\Delta_i S({\bf X}) \, = \, k_B \, Z({\bf X})
\end{equation}
where the quantity $Z({\bf X})$ is defined as
\begin{equation}
\label{Eq6}
Z({\bf X}) \, = \,  \ln \frac{ P({\bold X})}{ P^{(R)}(\widetilde{\bold X})}
\end{equation}
with the obvious assumption that  $\widetilde{{\bold X}}$ {\it is not} of zero measure.  This relation generalizes the traditional Gibbs-Shannon stochastic thermodynamics \cite{Nico} and, as such, it was at the origin of an impressive resurgence of interest in this field \cite{Rev1,Rev2,Jarz4}. The more so since $Z({\bf X})$ obeys the (detailed) fluctuation theorem, in the sense that
\begin{equation}
\label{Eq7}
\frac{P(\zeta)}{P(- \, \zeta)} \, = \, \frac{{\cal P}\Big\{\zeta \, < \, Z   \, \le \,  \zeta + d \zeta  \Big\}}{ {\cal P}\Big\{- \zeta \, < \, Z  \, \le \,  - \zeta + d \zeta  \Big\}}  \, = \, \exp(\zeta)
\end{equation}
where ${\cal P}\{E\}$ denotes the probability for the event $E$ to occur.  It is important to notice that the validity of this result goes far beyond the simple case of jump Markov processes. It doesn't rely on any physico-chemical property of the system, nor on the Markovian attribute of the underlying stochastic process, so long as the reverse process exists and remains bounded almost surely \cite{MF}.

The relation (\ref{Eq5}) was established by several authors, using different approaches \cite{Maes1,Crooks,dS1,dS2,dS3,dS4}.  We shall not go through the mathematical demonstration of this relation, nor discuss the physical justification of the underlying concepts (see \cite{Harris} and \cite{Rev2} for extensive reviews). The more so since this type of relation, with the associated fluctuation theorem, concerns a variety of systems subjected to different type of non-equilibrium constraints, such as controlled time dependent driving external forces and/or reservoirs, nicely illustrated  by Jarzynski using a Hamiltonian approach \cite{Jarz1}.  Instead, we will question the validity of (\ref{Eq5}) in reactive systems submitted to a given (time-independent) non-equilibrium constraint and modeled as a jump Markov process.

Since $\chi(t)$ is a Markov process, we can write  $P({\bf X})  =  P(X_0,  t_0) \times \prod_{i=1}^{n} P(X_{i},  t_i \, | \, X_{i - 1},  t_{i-1})$ and  $P^{(R)}(\widetilde{\bf X})  =  P^{(R)}(X_f,  t_0) \times \prod_{i=1}^{n} P^{(R)}(X_{n-i},  t_{i} \, | \, X_{n-i+1},  t_{i-1})$.
Inserting these relations into (\ref{Eq6}) and noticing that, by construction,  $P^{(R)}(X_f, t_0)  =  P(X_f, \, t_f)$, we find
\begin{equation}
\label{Eq8}
Z({\bf X})  =  \sum_{i=1}^{n} \,  \ln \frac{P(X_{i},  t_i \, | \, X_{i - 1},  t_{i-1})}{P^{(R)}(X_{i-1},  t_{n -i +1} \, | \, X_{i}, t_{n-i})}  +  \ln \, \frac{P(X_0, t_0)}{P(X_f, t_f)}
\end{equation}
For $n \gg 1$, we can use the relation (\ref{Eq4}) to obtain the main result
\begin{equation}
\label{Eq9}
Z({\bf X}) \, = \, \sum_{i=1}^{n} \,  \ln \frac{W(X_{i} \, | \, X_{i - 1})}{W(X_{i-1} \, | \, X_{i})} \, + \, \ln \, \frac{P(X_0, t_0)}{P(X_f, t_f)}
\end{equation}
This relation was first introduced by Lebowitz and Spohn \cite{Lebo} and later extended to the specific case of reactive processes by Gaspard \cite{Gasp}.  Since then, combined with (\ref{Eq5}), it constitutes the starting point of practically all path thermodynamics  formulations of reactive systems.

\section{Mathematical proofs}
We now establish two fundamental results:
First, we prove that the necessary condition for the validity of path thermodynamics is that there exists one, and only one, elementary reaction associated with each possible composition change of the system. Second, we prove that the class of reactive systems illustrated by the scheme (\ref{Eq1}) have sample paths that are time-reversible in the stationary regime, even though the system operates under non-equilibrium constraints.

Let us consider the specific case of a system involving $p \, \ge 2$ possible processes leading to the very same transition $X' \rightarrow X$. Denoting the corresponding transition rates by $W_\rho(X \, | \,  X')$, $\rho = 1, \, 2, \cdots p$, the relation (\ref{Eq4}) implies
\begin{equation}
\label{Eq10}
P(X, t + \Delta t \, | \, X', \, t)  =  \sum_{\rho = 1}^p \, W_\rho(X \, | \,  X') \, \Delta t \, + \, {\rm o}(\Delta t) \, , \,\, \forall \, X \, \ne \, X'
\end{equation}
which simply shows that assigning different probabilities to the very same event (here the jump $X' \rightarrow X$), is just absurd. We note that, from the strict mathematical point of view, is just a direct consequence of the Kolmogorov-Feller unicity theorem.

This observation is precisely at the heart of the problem.  In fact, in reactive systems the relation (\ref{Eq9}) is just a duplication of the relation (\ref{Eq8}) where $P(X_i, t_i \, | \, X_{i-1}, t_{i-1})$ is replaced by $W(X_{i} \, | \, X_{i - 1}) \, \Delta t_i$. Therefore, according to (\ref{Eq10}),  each $W(X_{i} \, | \, X_{i - 1})$ in (\ref{Eq9}) is necessarily the sum of all the transition rates associated with elementary reactions giving rise to the jump  $X_{i - 1} \rightarrow X_{i}$.
And, evidently, the same argument holds for the corresponding reverse reactions $X_{i} \rightarrow X_{i - 1}$.

It is important to notice that this fact is not a restriction at the level of the master equation formulation (\ref{Eq3}) where the contribution of each elementary process is treated separately, precisely because they appear as a sum \cite{Nico,Vdb1}.  The situation is different for the function $Z({\bf X})$ since it is defined as the sum of the logarithm of  transition rates, the argument of each logarithm function being the sum of all transition rates leading to the same jump.  Given that the entropy production of reactive systems depends on the characteristics of each individual reaction \cite{Dilip}, the relation (\ref{Eq5}) implies automatically that stochastic thermodynamics remains limited to systems where there exists one, and only one, elementary process associated to each possible jump.  This result completes the demonstration of our first statement.

Consider now the paradoxical issue illustrated by the reactive system (\ref{Eq1}) and let us assume that the associated stochastic process $\chi(t)$ possesses a stationary regime, i.e., the corresponding stationary probability distribution $P_s(X)$ exists.  Denoting the transition rates of this system by $\lambda_\rho(X) \equiv W_\rho(X + 1 \, | \, X)$ and $\mu_\rho(X) \equiv W_{-\rho}(X - 1\, | \, X)$, $\rho \in [1, \, \mathcal{R}]$, and inserting these expressions into the master equation (\ref{Eq3}), we can easily verify that the latter satisfies the principle of detailed balance at the stationary state, i.e.,
\begin{equation}
\label{Eq11}
\mu(X) \, P_{s}(X) \, = \,  \lambda(X - 1) \, P_{s}(X - 1)
\end{equation}
where $\lambda(X) = \sum_{\rho} \lambda_\rho(X)$ and $\mu(X) = \sum_{\rho} \mu_\rho(X)$.  This relation proves our second statement since we have a theorem stating that the necessary and sufficient condition for a stationary Markov process to be time-reversible is that it satisfies the principle of detailed balance (see, for example,  Section 6.3 of \cite{Gard} for details).
However, for the sake of completeness, we give here a slightly simpler proof.

We observe that, by definition, the sequence of states visited by $\chi(t)$ along the sample path ${\bf X}$ is arbitrary, with the exceptions that $ \{\chi(t_0) = X_0 \, ; \, \chi(t_f) = X_f \}$.  Being in a state $X_i$ at time $t_i$, the process may well remain there during $\Delta t_i = t_{i+1} - t_i$, so that $X_{i+1} = X_i$.  For the reactive system (\ref{Eq1}), there exist only two other possibilities: either $X_{i+1} = X_i + 1$ ($i \ne n$), in which case $W(X_{i+1} \, | \, X_{i}) = \lambda(X_i)$, or $X_{i+1} = X_i - 1$ ($X_i > 0$) and thus $W(X_{i+1} \, | \, X_{i}) = \mu(X_i)$.

In both cases, we can easily check that (\ref{Eq11}) can be written as
\begin{equation}
\label{Eq12}
W(X_{i} \, | \, X_{i - 1}) \, P_s(X_{i-1}) \, = \, W(X_{i-1} \, | \, X_{i}) \, P_s(X_{i})
\end{equation}
Consequently, the relation (\ref{Eq9}) reduces to
\begin{equation}
\label{Eq13}
Z({\bf X}) \, = \,   \ln \frac{P_s(X_f)}{P_s(X_0)} \, + \, \ln \, \frac{P(X_0, t_0)}{P(X_f, t_f)}
\end{equation}
which vanishes at the stationary regime, implying in turn that $P_s({\bold X}) = P_s^{(R)}(\widetilde{\bold X})$. This result completes the demonstration of our second statement.

One last point to be clarified concerns the Schnakenberg graph theory of reactive systems and its extension to jump Markov processes by Andrieux and Gaspard \cite{Gasp1}.  This theory is just an alternative perspective to the very same problem, that is, the study of the statistical properties of a non-equilibrium system from their sample paths.    And it suffers from exactly the same restrictions as those of the traditional path thermodynamics approach.  To be precise, if we follow the stochastic Schnakenberg's procedure by taking into account each elementary process, one by one, in ``multigraph'' situations, then we obtain an expression for the ``currents'' in non-equilibrium systems that indeed leads to the expected thermodynamic result, in average. And the very same property holds for the path entropy production obtained from the traditional path thermodynamics approach \cite{Gasp}, again in average.  The crucial point turns out to be the way the authors actually performed this average.  Rather than using sample paths of the system obtained by an appropriate succession of individual transition rates $W_\rho(X \, | \,  X')$ (see eq. (\ref{Eq10})) a close examination of their work shows that they actually use, quite wisely, the master equation based on the Kolmogorov equation (\ref{Eq12}) (eq. 45 in ref. [31] and eq. (20) in ref. [32]).  This strategy, however, has its own intrinsic limitations.  In particular, it cannot be used to establish a fluctuation theorem for non-equilibrium ``currents'', nor for the path entropy production.  For that, one way or another, we have to appeal to sample path properties of the system.

A ``sample path'' of a stochastic process $\chi(t)$ has a precise mathematical meaning.  It represents a succession of states assumed by $\chi(t)$ in the course of time.  These states are thus necessarily {\it measurable},   in the sense that a probability density can be associated to each of them individually.  In this respect it is quite instructive to recall that a fundamental theorem by Kolmogorov states that a stochastic process can be entirely characterized by the family of its sample paths if, and only if, the latter represent a succession of {\it measurable} states visited by the system in the course of time.  In particular, a transition rate associated to a reaction is a measurable quantity if, and only if, there exists no other elementary reaction leading to the same change.  The Kolmogorov relation (\ref{Eq10}) clearly proves that it is impossible to associate a (conditional) probability distribution to each individual transition rate $W_\rho(X \, | \,  X')$.  In other words, these transition rates {\it are not} separately measurable quantities so they cannot be used to define the stochastic process $\chi(t)$,  the latter being defined by the sum of all transition rates leading to the same jump.

Needless to say, we are entirely free to define a "path" any way we want. But then it is \emph{not} always possible to associate a stochastic process to an arbitrary constructed path.  This misinterpretation of the Kolmogorov's theorem is at the origin of the Gaspard-Andrieux's wrong result \cite{Gasp,Gasp1}.

We note that this statement is quite obvious from a physical point of view.  In fact, we have already shown that the sample paths of the reactive system (\ref{Eq1}) are time reversible at the stationary regime (i.e., $P_s({\bf X}) = P_s(\widetilde{\bf X})$) so extracting an {\it irreversible} property from {\it strictly reversible trajectories} is just a fool's errand.  In particular, no matter how we define the current or the entropy production, in terms of elementary transition rates of the reactive system (\ref{Eq1}), they cannot obey the fluctuation theorem.  For the sake of completeness, we nevertheless present below the mathematical proof.

The authors consider explicitly the case of the Schl{\"o}gl model ($\mathcal{R} = 2$, in (\ref{Eq1})).  This model involves two elementary reactions leading to the jump $X_i \rightarrow X_i + 1$ and two others leading to the reverse jump  $X_{i} \rightarrow X_i - 1$. Let us denote the corresponding transition rates by $W_{\rho_i}(X_i + 1 \, | \, X_i)$ and  $W_{- \rho_{i}}(X_i - 1 \, | \, X_i)$, respectively, where the subscript $\rho_i = 1$ or $2$.  Following the authors, one obtains at the stationary regime (cf. eq. (\ref{Eq5})),
\begin{equation}
\label{Eq14}
Z({\bf X})  =  \Delta_i S({\bf X}) /k_B   =  \sum_{i=1}^{n} \,  \ln \frac{ W_{\rho_i}(X_{i} | X_{i-1})  }{ W_{- \rho_i} (X_{i - 1} | X_{i}) }  +   \ln \, \frac{P_s(X_0)}{P_s(X_f)}
\end{equation}
where the values assigned to $\rho_i$ (1 or 2) determine the precise structure of the path leading to $X_0 \rightarrow X_f$ and its reverse. As stated above, the authors showed that the average $\langle \Delta_i S({\bf X}) \rangle$ is  equivalent to the expected thermodynamic result.

Let us now consider the probability density $P(\zeta)$ associated to $Z({\bf X}$), defined as
\begin{equation}
\label{Eq15}
{\cal P}\Big\{\zeta  \, < \, Z({\bf X}) \,\, \le \,\,  \zeta  +  d \zeta \Big\} \, = \, P(\zeta) \, d \zeta
\end{equation}
Using the fundamental relation $P_s({\bold X}) = P_s^{(R)}(\widetilde{\bold X})$  and noticing that the change of variables ${\bold X} \rightarrow \widetilde{\bold X}$ transforms  $P({\bold X})$ to $P^{(R)}(\widetilde{\bold X})$, $P_s(X_0)$ to $P_s(X_f)$,  $W_{\rho_i}(X_i + 1 \, | \, X_i)$ to  $W_{- \rho_{i}}(X_i - 1 \, | \, X_i)$, and thus $Z({\bf X})$ to $-  \,Z({\bf X})$, we readily find that at the stationary regime
\begin{eqnarray}
\label{Eq16}
P(\zeta)  \! & = & \!  \sum_{\bf X} \, \delta^{Kr}_{\zeta, \, Z({\bf X})} \,\,  P_s({\bold X}) \, = \, \sum_{\widetilde{\bf X}} \, \delta^{Kr}_{\zeta, \, Z(\widetilde{\bf X})} \,\, P_s^{(R)}(\widetilde{\bold X}) \nonumber \\
& = &  \sum_{\bf X}  \, \delta^{Kr}_{\zeta, \, - \, Z({\bf X})} \,\, P_s({\bold X}) \,  =  \, P(- \, \zeta)\end{eqnarray}
Consequently, $P(\zeta) = P(- \, \zeta)$ which contradicts (\ref{Eq7}) and thus proves that the entropy production defined by (\ref{Eq14}) does not obey the fluctuation theorem.  Clearly, for these types of reactive systems using the Schnakenberg graph theory formulation does not correct the flaw that exists with path thermodynamics.

\section{Concluding remarks}
Obviously, the core of the problem lies in the fact that the specificities of elementary reactions cannot be deduced from the sample paths of the corresponding reactive system in "multigraph" situations. One potential way out of this difficulty is to supplement the stochastic formulation with a sort of "counting parameter" which is used to identify the actual elementary process responsible for each transition.  This strategy was proposed by Gaspard et al. for the stochastic Schnakenberg formulation of reactive systems \cite{Gasp2}, as well as for the so-called "quantum dots" problem \cite{Gasp3} (an entirely different issue).

The legitimate question is obviously the precise meaning of this "counting parameter".  One way to implement this parameter is through Gillespie's numerical simulation method.  Being at a given state, we pick a random number to choose the next jump. If there exist several elementary processes associated with that jump, then we pick a second random number to decide which reaction is actually responsible for that jump. This method allows us to construct a random path thus justifying the expression $\Delta_i S({\bf X})$ of path entropy production proposed by Gaspard and Andrieux for the Schl{\"o}gl model (\ref{Eq14}).  No doubt that upon repeating this procedure  as many times as necessary, we will find that, in average, $\Delta_i S({\bf X})$ approaches the expected thermodynamic behavior.  Why? Simply because generating the extra random numbers between successive jumps does not modify the dynamics of the system, which therefore remains identical to  Kolmogorov's original formulation.  The process of averaging by repeating over and over again the same sequence of numerical experiments, is thus identical to take the average at the level of master equation (see the discussions after eq. (\ref{Eq13})). The drawback, however, is that the resulting path generated in this way can no longer be associated with the sample path of the system.  As shown above, the path entropy production defined in this way does not obey the fluctuation theorem.

Another possibility, is to incorporate this counting parameter to the very definition of the stochastic process, a procedure that needs obviously to be define in a precise way.  However, it is important to keep in mind that this alternative procedure changes profoundly the statistical nature of the problem because the system is provided with some information that it does not actually possesses.  Determining thermodynamic properties of a system under this modified condition is, at minimum, a problematic endeavor that requires evidently more profound investigations.

One way to decide once and for all this controversial issue is to appeal to reactive Boltzmann simulations. Here, the elastic and reactive collisions between the particles (usually, hard spheres) obey precise microscopic rules, and the evolution of the composition of the system is entirely determined by the knowledge of the activation energies of the reactive processes, thus excluding the introduction of any extra parameters or hidden variables.
%We have undertaken such simulations, which are quite CPU intensive, and thus far the observed results have confirmed all of our previous expectations. The final article describing this study is in preparation.

One last perspective needs to be mentioned.  We know that in the limit of large system size, a jump Markov process converges to a diffusion process whose associated probability obeys a Fokker-Planck equation (the famous Kurtz theorem \cite{Kurtz}). What about the validly of path thermodynamics in this case?  As we will soon report elsewhere, yet again a number of path thermodynamics predictions turn out to be simply wrong.

\section*{Acknowledgments}

We acknowledge the contributions of Professors Gregoire Nicolis and Christian Van den Broeck with whom the  authors have over the years many insightful and productive discussions.


\begin{thebibliography}{99}

\bibitem{Nico}	Luo Jiu-Li, C. Van den Broeck, and G. Nicolis, Z. Phys. B: Condens. Matter {\bf 56}, 165 (1984).

\bibitem{Rev1} C. Bustamante, J. Liphardt, F. Ritort, Phys. Today {\bf 58} (7) 43 (2005) ; D. J. Evans, G. Morris, Statistical Mechanics of Nonequilibrium Liquids, second ed., Cambridge University Press, (2008).

\bibitem{Harris}  R. J. Harris	and	G. M. Schutz,	J. Stat. Mech., {\bf 07}, P07020 (2007).

\bibitem{Rev2} M. Esposito, U. Harbola, S. Mukamel, Rev. Modern Phys. {\bf 81}, 1665 (2009) ; X.-J. Zhang, H. Qian, M. Qian, Phys. Rep. {\bf 510}, 1 (2012) ; H. Ge, M. Qian, H. Qian, Phys. Rep. {\bf 510}, 87 (2012) ; U. Seifert, Rep. Progr. Phys. {\bf 75} 126001 (2012).

\bibitem{Jarz4} C. Jarzynski, Annu. Rev. Condens. Matter Phys. {\bf 2}, 329 (2011) ; H. Ge and H. Qian, Phys. Rev. E {\bf 81}, 051133 (2010).

\bibitem{MF} M. Malek Mansour and F. Baras, Chaos {\bf 27}, 104609 (2017).

\bibitem{Vidal} C. Vidal and A. Pacault (Eds), {\it Nonlinear Phenomena in Chemical Dynamics}, Springer-Verlag, Berlin (1981).

\bibitem{Adv} F. Baras and M. Malek Mansour, Adv. Chem. Phys. {\bf 100}, 393 (1977).

\bibitem{Gil1}   D. T. Gillespie, J. Comput. Phys. {\bf 22}, 403 (1976) ; J. Phys. Chem. {\bf 81}, 2340
(1977).

\bibitem{Dilip} D. Kondepudi, {\it Introduction to Modern Thermodynamics}, Wiley (2008).

\bibitem{Kawai} R. Kawai, J. M. R. Parrondo, and C. Van den Broeck, Phys. Rev. Lett. {\bf 98}, 080602 (2007).

\bibitem{Vdb1} M. Esposito and C. Van den Broeck, Phys. Rev. Lett. {\bf 104}, 090601 (2010) ;  Phys. Rev. E {\bf 82}, 011143 (2010).

\bibitem{Russe} D. Seleznev, G. A. Zhernokleeva and L. M. Martyushe, JETP Letters {\bf 102}, 557 (2015).

\bibitem{Kurch} J. Kurchan, J. Phys. A {\bf 31}, 3719 (1998).

\bibitem{DDK} Y. De Decker, A. Garcia Cantu Ros and G. Nicolis, Euro. Phys. J. {\bf 224}, 947 (2015) ; Y. De Decker, J-F. Derivaux and G. Nicolis, Phys. Rev. E {\bf 93},  042127 (2016).

\bibitem{NicPrig} G. Nicolis, and I. Prigogine,
\textit{Self-Organization in Nonequilibrium Systems},  Wiley, (1977).

\bibitem{Epst} I. Epstein and J. Pojman, \textit{An Introduction to Nonlinear Chemical Dynamics},  Oxford University Press, (1998).

\bibitem{DegMaz} S. R. de Groot and P. Mazur, {\it Non-equilibrium Thermodynamics}, Dover, New York, (1984).

\bibitem{Gard}	 C.W. Gardiner, {\it Handbook of Stochastic Methods}, Springer-Verlag (2009).

\bibitem{VK}	N. G. Van Kampen, {\it Stochastic Processes in Physics and
Chemistry}, North-Holland, Amsterdam (1983).

\bibitem{Kolmo} See for example, B. {\O}ksendal, {\it Stochastic Differential Equations: An Introduction with Applications}. Springer (2003).

\bibitem{Soong} T. Soong, \textit{Random Differential Equations in Science and Engineering}, Academic Press, (1973).

\bibitem{Maes1} C. Maes, J. Stat. Phys., {\bf 95}, 367 (1999) ; C. Maes and K. Netocn\'y, J. Stat. Phys., {\bf 110}, 269 (2003) ;  C. Maes and F. Redig,  J. stat. Phys., {\bf 101}, 3 (2000).

\bibitem{Crooks} G. E. Crooks, Phys. Rev. E {\bf 60}, 2721 (1999) ;  Phys. Rev. E {\bf 61}, 2361 (2000).

\bibitem{dS1} U. Seifert, Phys. Rev. Lett. {\bf 95}, 040602 (2005) ; Eur. Phys. J. B {\bf 64}, 423 (2008).

\bibitem{dS2} C. Jarzynski, Phys. Rev. E {\bf 73}, 046105 (2006).

\bibitem{dS3} M. Esposito, U. Harbola, and S. Mukamel, Phys. Rev. E {\bf 76}, 031132 (2007).

\bibitem{dS4}  R. Kawai, J. M. R. Parrondo, and C. Van den Broeck, Phys. Rev. Lett. {\bf 98}, 080602 (2007) ; J. M. R. Parrondo, C. Van den Broeck, and R. Kawai, New J. Phys. {\bf 11}, 073008 (2009).

\bibitem{Jarz1}	 C. Jarzynski, J. Stat. Phys. {\bf 98}, 77 (2000).

\bibitem{Lebo}	 J. L. Lebowitz and H. Spohn, J. Stat. Phys., {\bf  95}, 333 (1999).

\bibitem{Gasp}	P. Gaspard, J. Chem. Phys. {\bf 120}, 8898 (2004).

\bibitem{Gasp1}	 D. Andrieux and P. Gaspard, J. Chem. Phys. {\bf 121}, 6167 (2004) ; ibid  {\bf 128}, 154506 (2008).

\bibitem{Gasp2}	 D. Andrieux and P. Gaspard, J. Stat. Phys. {\bf 127}, 107 (2007).

\bibitem{Gasp3}	 G. Bulnes Cuetara, M. Esposito, and P. Gaspard, Phys. Rev. {\bf 84}, 165114 (2011).

\bibitem{Kurtz}	T. G. Kurtz, Math. Progr. Stud. 5, {\bf 67} (1976).

\end{thebibliography}
\end{document}